\documentclass[aps,pra,groupedaddress,twocolumn]{revtex4-1}

\usepackage{graphicx}
\usepackage{dcolumn}
\usepackage{amssymb}
\usepackage{amsmath}
\usepackage{bbold}
\usepackage{bm}
\usepackage{color}
\usepackage{physics}
\usepackage{chemformula}

\usepackage{soul}

\usepackage[colorlinks,urlcolor=blue,citecolor=blue,linkcolor=blue]{hyperref}
\usepackage{comment}
\usepackage{cleveref}
\usepackage{mathtools}
\usepackage{mathrsfs}

\newcommand{\eb}{\varepsilon_B}

\renewcommand{\k}{\mathbf{k}}
\newcommand{\ek}{\epsilon_\k}

\newcommand{\p}{\mathbf{p}}
\newcommand{\ep}{\epsilon_\p}
\newcommand{\q}{\mathbf{q}}

\newcommand{\kone}{\mathbf{k}_1}
\newcommand{\ktwo}{\mathbf{k}_2}
\newcommand{\kthree}{\mathbf{k}_3}

\newcommand{\Xh}{\hat{X}}

\makeatletter
\newcommand{\subalign}[1]{%
  \vcenter{%
    \Let@ \restore@math@cr \default@tag
    \baselineskip\fontdimen10 \scriptfont\tw@
    \advance\baselineskip\fontdimen12 \scriptfont\tw@
    \lineskip\thr@@\fontdimen8 \scriptfont\thr@@
    \lineskiplimit\lineskip
    \ialign{\hfil$\m@th\scriptstyle##$&$\m@th\scriptstyle{}##$\hfil\crcr
      #1\crcr
    }%
  }%
}
\makeatother

\begin{document}

\title{Variational approach to the two-dimensional Bose polaron}

\author{Yasufumi Nakano}
\affiliation{School of Physics and Astronomy, Monash University, Victoria 3800, Australia}
\affiliation{ARC Centre of Excellence in Future Low-Energy Electronics Technologies, Monash University, Victoria 3800, Australia}

\author{Meera M. Parish}
\affiliation{School of Physics and Astronomy, Monash University, Victoria 3800, Australia}
\affiliation{ARC Centre of Excellence in Future Low-Energy Electronics Technologies, Monash University, Victoria 3800, Australia}

\author{Jesper Levinsen}
\affiliation{School of Physics and Astronomy, Monash University, Victoria 3800, Australia}
\affiliation{ARC Centre of Excellence in Future Low-Energy Electronics Technologies, Monash University, Victoria 3800, Australia}

\date{\today}

\begin{abstract}
An impurity particle interacting with a Bose-Einstein condensate (BEC) leads to the formation of a quasiparticle known as the Bose polaron. We investigate the properties of the two-dimensional Bose polaron, applying a variational ansatz that contains up to three Bogoliubov excitations of the BEC. Similar to its three-dimensional counterpart, we observe the existence of two quasiparticle branches, namely the attractive and the repulsive polarons, at different coupling strengths. We find that their energies agree well with recent quantum Monte Carlo calculations. In particular, we observe that the inclusion of three excitations is crucial to capture the attractive polaron energy towards the regime of strong attraction, where the quasiparticle properties are dominated by few-body correlations. We also calculate the attractive polaron effective mass and residue, where we find significant differences between considering a weakly interacting Bose medium and taking the non-interacting limit, signalling enhanced impurity dressing by excitations in the latter case. 
By contrast, the spectral weight of the metastable repulsive polaron is largely insensitive to the interactions in the BEC and the number of Bogoliubov excitations. Our model may be experimentally realized in dilute atomic vapors and atomically thin semiconductors.
\end{abstract}

\maketitle

\section{Introduction}
The interaction of an impurity particle with a quantum mechanical medium, the so-called quantum impurity problem, plays a key role in understanding quantum many-body physics. One example is the polaron, originally proposed by Landau and Peker \cite{Landau1948}, that is formed by a mobile electron dressed by a cloud of virtual phonons in an ionic crystal. The concept of the polaron has since been widely extended: In the case of ultracold atomic gases, the presence of an impurity atom in a Bose-Einstein condensate (BEC) gives rise to the formation of the \textit{Bose polaron}, where the impurity becomes dressed by Bogoliubov excitations of the BEC~\cite{Tempere2009}. Here, one can take advantage of a magnetically tunable Feshbach resonance that enables the precise control of the coupling strength between the impurity and the medium, thus allowing one to investigate the polaron quasiparticle properties, as recently demonstrated in experiments~\cite{Jorgensen2016,Hu2016,Yan2020}. This poses an intriguing test for theoretical modelling, and indeed the Bose polaron has been studied using a variety of theoretical methods, ranging from field-theoretical diagrammatic approaches~\cite{Rath2013,Guenther2018} and variational methods~\cite{Li2014,Levinsen2015,Shchadilova2016,Yoshida2018PRX,Drescher2019,Field2020,Christianen2022}, to quantum Monte Carlo (QMC) methods \cite{Ardila2015,Ardila2016}, renormalization group theory \cite{Grusdt2017}, and a high-temperature virial expansion \cite{Sun2017}.

A particularly interesting aspect of the Bose polaron is its relationship to few-body bound clusters involving the impurity and several bosons from the medium. Unlike the widely studied Fermi polaron~\cite{Massignan2013,Scazza2022,Parish2023}---an impurity immersed in a fermionic medium---the Bose polaron does not feature any transitions in its ground state, and thus the polaron can continuously change its character from a weakly dressed impurity to a state that has strong few-body correlations. In a three-dimensional (3D) system, where three particles can bind to form Efimov trimers~\cite{Naidon2017}, it was shown in Ref.~\cite{Yoshida2018PRX} that, in the case of the equal-mass system, the behavior of the Bose polaron at strong interactions is characterized by the size of the ground-state trimer and that the polaron energy is a universal function of the Efimov three-body parameter~\cite{Braaten2006} for sufficiently low boson densities.
The existence of Efimov trimers depends strongly on dimensionality, being absent in the ideal two-dimensional (2D) case~\cite{Bruch1979,Lim1980} and suppressed under realistic transverse confinements~\cite{Levinsen2014b}.
This means that, unlike the 3D Bose polaron, the properties of the 2D Bose polaron do not depend on the introduction of a three-body parameter~\cite{Berninger2011}, and are instead fully characterized by the interparticle spacing of the medium relative to the characteristic length scales of the interactions such as the impurity-boson and boson-boson scattering lengths. On the other hand, the system supports the formation of two- and three-body bound states at any scattering length~\cite{Bruch1979,Pricoupenko2010,Levinsen2015review}, unlike the 3D case, and one might therefore wonder how these bound states affect the polaron.
The 2D Bose polaron has been previously investigated using the  Fr{\"o}hlich model \cite{Casteels2012,Pastukhov2018b}, a variational approach for light-matter coupled polarons~\cite{Levinsen2019}, a mean-field approach \cite{Hryhorchak2020}, a non-self-consistent $T$-matrix approximation approach \cite{Castillo2023}, QMC \cite{Ardila2020}, and renormalization group theory \cite{Isaule2021}. However, the detailed connection between few- and many-body physics for the 2D Bose polaron in an ultracold atomic gas remains unexplored.

In this paper, we investigate the properties of the 2D Bose polaron by applying a variational ansatz that contains up to three Bogoliubov excitations of the BEC. Similarly to the 3D Bose polaron~\cite{Scazza2022}, we observe two distinct quasiparticle branches: the attractive and repulsive polarons characterized by negative and positive energies, respectively. For a large range of weak to intermediate attractive interaction strengths, we observe that including three excitations of the BEC as well as interactions within the BEC is necessary to accurately reproduce QMC results~\cite{Ardila2020} for the attractive polaron energy. Furthermore, this level of approximation gives qualitative agreement also for the quasiparticle residue and effective mass of the attractive polaron. Including additional excitations of the medium would likely substantially improve the agreement in the strongly interacting regime where the interparticle spacing is comparable to the size of the impurity-boson bound state. For the repulsive polaron, we find that the energy is well reproduced already with a single excitation; however the residue behaves very differently to that in the QMC, indicating that the metastability of the repulsive polaron plays an important role and requires further study.

This paper is organized as follows. In Sec. \ref{sec:model and variational approach} we describe our model of an impurity in a 2D BEC and outline our variational approach. In Sec. \ref{sec:polaron properties} we discuss the polaron properties including our results for the energy, residue, and effective mass, and compare these with results from QMC~\cite{Ardila2020}. We conclude in Sec. \ref{sec:conclusions}.

\section{Model and Variational Approach}
\label{sec:model and variational approach}

\subsection{Model}
\label{subsec: model}

We consider an impurity particle immersed in a weakly interacting uniform two-dimensional Bose gas, where we focus on the zero-temperature case such that the system features a BEC~\footnote{Even in the case of a small but finite temperature, where there is no true long-range order in the infinite system, there can still be a well-defined condensate in a finite-sized or a harmonically-trapped system}.
We model the system using a two-channel description of a Feshbach resonance \cite{Timmermanns1999}, similarly to how the Bose polaron was described in the three-dimensional case in Refs.~\cite{Levinsen2015,Yoshida2018PRX}. Measuring the energy with respect to that of the BEC, the Hamiltonian is given by (setting $\hbar$ and the area to 1):
\begin{align}
   &\hat{H} =\sum_{\k}\left[E_{\k}\beta^{\dagger}_{\k}\beta_{\k} + \ek c^{\dagger}_{\k}c_{\k} + (\ek^{d}+\nu_{0})d^{\dagger}_{\k}d_{\k}\right] \notag \\
   + &g\sqrt{n_{0}}\sum_{\k}\left(d^{\dagger}_{\k}c_{\k} + h.c. \right) + g\sum_{\k,\q}\left(d^{\dagger}_{\q}c_{\q-\k}b_{\k} + h.c.\right) \label{eq:Hamiltonian}.
\end{align}
Here, $b^{\dagger}_{\k}$ and $c^{\dagger}_{\k}$ correspond to the creation operators of a boson and the impurity, respectively, with momentum ${\k}$. $\ek = \frac{k^2}{2m}$ is the corresponding single-particle energy dispersions, where we assume that the impurity and bosons have equal masses $m$. $d^{\dagger}_{\k}$ corresponds to the creation operator of a closed-channel molecule, which is formed from the impurity and a boson when they interact, and we have $\ek^d = \frac{k^2}{4m}$. The bare detuning $\nu_{0}$ corresponds to the energy of the closed channel relative to the open impurity-boson channel.

In writing the Hamiltonian, we have applied the Bogoliubov theory of the weakly interacting Bose gas~\cite{FetterBook}, where we have $n_0a_{\rm B}^2\ll1$ in terms of the (positive) 2D boson-boson scattering length $a_{\text{B}}$ and the condensate density $n_{0}$. The Bogoliubov dispersion is given by
\begin{equation}
    E_{\k} = \sqrt{\ek(\ek + 2\mu)},
\end{equation}
with the chemical potential $\mu = \frac{4\pi n_{0}/m}{\ln(1/n_{0}a^{2}_{\text{B}})}$~\cite{Schick1971}. 
The bare boson operator $b_{\k}$ is related to the Bogoliubov operator via the Bogoliubov transformation:
\begin{equation}
    b_{\k} = u_{\k}\beta_{\k}-v_{\k}\beta^{\dagger}_{-\k} \label{eq:Bogoliubov transformation},
\end{equation}
with the coherence factors:
\begin{align}
    u_{\k} &= \sqrt{\frac{1}{2}\left(\frac{\ek+\mu}{E_{\k}}+1\right)},\quad v_{\k} &= \sqrt{\frac{1}{2}\left(\frac{\ek+\mu}{E_{\k}}-1\right)}.
\end{align}

The presence of the impurity adds the scattering length $a_{\text{2D}}$ to the system, which characterizes the interactions between the impurity and the medium. We assume that both the boson-boson and impurity-boson interactions are contact, $s$-wave interactions. This assumption is valid since the average interparticle distance and the thermal wavelength far exceed the range of the underlying van der Waals interactions. In applying the Bogoliubov theory of the weakly interacting Bose gas, we have already implicitly assumed that the boson-boson interactions are short-ranged.

To be specific, the second line in Eq.~\eqref{eq:Hamiltonian} describes the interactions between the impurity and a boson with coupling constant $g$, which proceed via the formation of a closed-channel molecule. The coupling constant $g$ and the bare detuning $\nu_{0}$ can be related to the most general form of the low-energy $s$-wave 2D scattering amplitude~\cite{landau2013quantum}:
\begin{equation}
    f_{\text{2D}}(k) = \frac{4\pi}{-\ln(k^{2}a^{2}_{\text{2D}}) + R^{2}_{\text{2D}}k^{2} + i\pi}, \label{eq:scattering amplitude}
\end{equation}
where $R_{\text{2D}}$ is the 2D range parameter. Calculating the scattering amplitude within the two-channel model~\eqref{eq:Hamiltonian} and carrying out the renormalization procedure~\cite{Levinsen2015review}, we obtain the scattering amplitude and range parameter~\cite{Kirk2017}:
\begin{equation}
    a_{\text{2D}} = \frac{1}{\Lambda}e^{\frac{2\pi\nu_{0}}{mg^2}}, \quad R_{\text{2D}} = \sqrt{\frac{4\pi}{m^2g^2}},
\end{equation}
where $\Lambda$ is an ultraviolet momentum cut-off on the relative impurity-boson momentum. For the results presented in this paper, we take $R_{\text{2D}} \rightarrow 0$, which corresponds to taking the limit of $g \rightarrow \infty$ and $\Lambda \rightarrow \infty$ while adjusting $\nu_0$ to keep $a_\text{2D}$ finite. This limit is well defined and corresponds to considering a single-channel model. We note that, although the interacting part of the Hamiltonian in Eq.~\eqref{eq:Hamiltonian} is written in terms of the bare boson operator, it is always related to the Bogoliubov operator via the Bogoliubov transformation in Eq.~\eqref{eq:Bogoliubov transformation}.

Importantly, the impurity-boson interaction always features a bound state in our two-dimensional setting. This so-called dimer state corresponds to the pole of the scattering amplitude, i.e., $f^{-1}_{\text{2D}} = 0$. In the limit of $R_{\text{2D}} \rightarrow 0$, the dimer binding energy $\eb$ takes the form: $\eb = 1/ma^{2}_{\text{2D}}$. Likewise, there exists a three-body bound state consisting of the impurity and two bosons~\cite{Bruch1979}, with energy $E_{\text{T}} = -2.39\eb$  in the limit $R_{\text{2D}} \rightarrow 0$ and $a_{\mathrm B}\to0$. Larger multi-body clusters are also predicted to exist in this system~\cite{Guijarro2020}.

In the experiment, the quasi-2D geometry may be achieved by applying a harmonic confining potential to a 3D atomic gas along the direction perpendicular to the 2D plane~\cite{Petrov2001}. In this case, the scattering properties of the 3D system can be mapped to the quasi-2D system~\cite{Kirk2017}. The 2D limit is achieved when the size of the two-body bound state, average interparticle spacing, and thermal wavelength are all much larger than the confinement length.

\subsection{Variational Ansatz}
\label{subsec: variational approach}
To investigate the properties of the polaron, we consider a variational state that describes the impurity and its dressing by up to three Bogoliubov excitations of the medium. Considering for simplicity a variational state of zero total momentum, this takes the form:
\begin{equation}
    \ket{\Psi} = \ket{\psi_0} + \ket{\psi_1} + \ket{\psi_2} + \ket{\psi_3},
    \label{eq:variational wavefunction}
\end{equation}
where $\ket{\psi_{N}}$ denotes a state with $N$ Bogoliubov excitations:
\begin{align}
    \ket{\psi_0} &= \alpha_{0}c^{\dagger}_{0}\ket{\Phi}, \notag \\
    \ket{\psi_1} &= \left(\sum_{\k}\alpha_{\k}c^{\dagger}_{-\k}\beta^{\dagger}_{\k} + \gamma_{0}d^{\dagger}_{0} \right)\ket{\Phi}, \notag \\
    \ket{\psi_2} &= \left(\frac{1}{2}\sum_{\kone\ktwo}\alpha_{\kone\ktwo}c^{\dagger}_{-\kone-\ktwo}\beta^{\dagger}_{\kone}\beta^{\dagger}_{\ktwo} + \sum_{\k}\gamma_{\k}d^{\dagger}_{-\k}\beta^{\dagger}_{\k} \right)\ket{\Phi}, \notag \\ 
    \ket{\psi_3} &= \left( \frac{1}{6}\sum_{\kone\ktwo\kthree}\alpha_{\kone\ktwo\kthree}c^{\dagger}_{-\kone-\ktwo-\kthree}\beta^{\dagger}_{\kone}\beta^{\dagger}_{\ktwo}\beta^{\dagger}_{\kthree} \right. \notag \\
    &\quad \left. + \frac{1}{2}\sum_{\kone\ktwo}\gamma_{\kone\ktwo}d^{\dagger}_{-\kone-\ktwo}\beta^{\dagger}_{\kone}\beta^{\dagger}_{\ktwo} \right)\ket{\Phi}.
    \label{eq:variational wavefunction 2}
\end{align}
Here, $\ket{\Phi}$ corresponds to the weakly interacting 2D BEC. The first line of Eq.~\eqref{eq:variational wavefunction 2} describes the bare impurity, while the second line is a superposition of the impurity dressed by a single Bogoliubov excitation, and a term where the impurity has bound a particle from the BEC to form a closed-channel dimer. The third (fourth) line describes a superposition of the impurity dressed by two (three) Bogoliubov excitations and a closed-channel dimer dressed by a single (two) Bogoliubov excitation(s). In this paper, we refer to calculations with the variational state up to $N = 1$ boson as two-body correlations, $N = 2$ as three-body correlations, and $N = 3$ as four-body correlations. $\alpha_{0}$, $\alpha_{\k}$, $\alpha_{\kone\ktwo}$, $\alpha_{\kone\ktwo\kthree}$, $\gamma_{0}$, $\gamma_{\k}$, and $\gamma_{\kone\ktwo}$ are variational parameters that are normalized according to $\bra{\Psi}\ket{\Psi} = 1$. These are determined by considering the stationary condition $\bra{\partial\Psi}(\hat{H}-E)\ket{\Psi} = 0$, where the derivative is taken with respect to each variational parameter, yielding a set of coupled linear integral equations for the variational parameters. These equations are identical to those that were originally derived in Refs.~\cite{Levinsen2015,Yoshida2018PRX} in the three-dimensional case, and therefore their explicit form is relegated to Appendix \ref{appx: couped integral equations}.

In the case of the 3D Bose polaron, the variational ansatz in Eq.~\eqref{eq:variational wavefunction} was shown in Ref.~\cite{Levinsen2015} to capture the physics associated with the exotic three-body Efimov bound states~\cite{Efimov1970}. Although Efimov trimers do not exist in the 2D system, we still have a trimer bound state at any scattering length, where the trimer binding energy can be calculated by taking the few-body limit, $n_{0} \rightarrow 0$, of the coupled equations in Eq.~\eqref{eq:coupled integral equations}.

\subsection{Impurity Spectral Function}
\label{subsec:impurity spectral function}
Of particular interest in experiments is the spectral response of the impurity in the medium. To form the polaron, a radio frequency (RF) pulse is used to transfer the impurity from a non-interacting (auxilliary) state into the interacting state. The transition probability is proportional to the impurity spectral function:
\begin{equation}
    A(\omega) = \sum_{j}\big|\bra{\Psi_{0}}\ket{\phi_{j}}\big|^{2}\delta(\omega-E_{j}),
    \label{eq:transition probability}
\end{equation}
where $\ket{\Psi_{0}}$ denotes the non-interacting polaron state: $\ket{\Psi_{0}} = c^{\dagger}_{0}\ket{\Phi}$. Here, $\ket{\phi_{j}}$ denotes the eigenstates of the Hamiltonian in Eq.~\eqref{eq:Hamiltonian} truncated to the Hilbert space described by the variational ansatz in Eq.~\eqref{eq:variational wavefunction}, and $E_{j}$ denotes their corresponding eigenvalues.  
To incorporate a broadening of the spectrum due to the finite duration of the RF pulse in experiment, we convolve the spectral function with a Gaussian of Fourier width $\sigma_{\text{rf}}$:
\begin{equation}
    I_{0}(\omega) = \int d\omega^{\prime} A(\omega-\omega^{\prime}) \frac{1}{\sqrt{2\pi}\sigma_{\text{rf}}} e^{-\omega^{\prime 2}/2\sigma^{2}_{\text{rf}}}.
    \label{eq:impurity specrtral function 1}
\end{equation}
Inserting Eq.~\eqref{eq:transition probability} in Eq.~\eqref{eq:impurity specrtral function 1}, we obtain the broadened impurity spectral function:
\begin{equation}
    I_{0}(\omega) = \sum_{j}\big|\bra{\Psi_{0}}\ket{\phi_{j}}\big|^{2} \frac{1}{\sqrt{2\pi}\sigma_{\text{rf}}} e^{-(\omega-E_{j})^{2}/2\sigma^{2}_{\text{rf}}}.
\end{equation}
As we shall see, this spectral function allows us to capture the peaks of the attractive and repulsive polarons as well as the continuum of states across a range of coupling strengths. In the present work, we use a Fourier broadening $\sigma_{\text{rf}}= 0.4 n_0/m$, which is comparable to the Aarhus experiment of the 3D polaron \cite{Jorgensen2016} and gives reasonable numerical convergence in the spectral response similarly to the 3D case \cite{Field2020}.

\section{Polaron Properties}
\label{sec:polaron properties}
We now discuss the quasiparticle properties of the 2D Bose polaron. In our variational approach, we obtain the energy spectrum by expressing the coupled equations in Eq.~\eqref{eq:coupled integral equations} as an eigenvalue equation and numerically evaluating the eigenvalues and eigenvectors on a discrete grid~\cite{numericalrecipes}. This also allows us to obtain the quasiparticle residue $Z$, i.e., the squared overlap between interacting and non-interacting states. To evaluate the polaron effective mass $m^*$, we extend our variational ansatz to the Bose polaron with a finite momentum (see Appendix \ref{appx: effective mass}). 

Note that the variational ansatz with one Bogoliubov excitation (i.e., two-body correlations) is formally equivalent to the non-self-consistent $T$-matrix approximation (NSCT). Indeed a very recent work~\cite{Castillo2023} applied NSCT to the 2D Bose polaron and obtained results similar to our lowest-order ansatz.
However, our full variational ansatz includes three- and four-body correlations as well as  bound states that go beyond this diagrammatic approach.

In the following, we compare our results with those from QMC~\cite{Ardila2020} and perturbation theory~\cite{Pastukhov2018b}. In particular, the energy, residue, and effective mass within the weak-coupling perturbative limit $|\ln(\sqrt{n_0}a_{\mathrm 2D})|\gg1$ take the forms~\cite{Pastukhov2018b}:
\begin{equation}
    \frac{E}{n_{0}/m} = -\frac{2\pi}{\ln{(\sqrt{n_0}a_\text{2D})}},
\end{equation}
\begin{equation}
    Z = \left[1 - \frac{1}{2}\frac{\ln{(\sqrt{n_0}a_\text{B})}}{\ln^2{(\sqrt{n_0}a_\text{2D})}}\right]^{-1},
\end{equation}
\begin{equation}
    \frac{m}{m^{\ast}} = 1 + \frac{1}{4}\frac{\ln{(\sqrt{n_0} a_\text{B})}}{\ln^2{(\sqrt{n_0}a_\text{2D})}},
\end{equation}
respectively.

\subsection{Polaron Energy}
\label{subsec:polaron energy}
\begin{figure}[htp]
    \centering
    \begin{minipage}{0.48\textwidth}
    \centering
    \includegraphics[width=\textwidth]{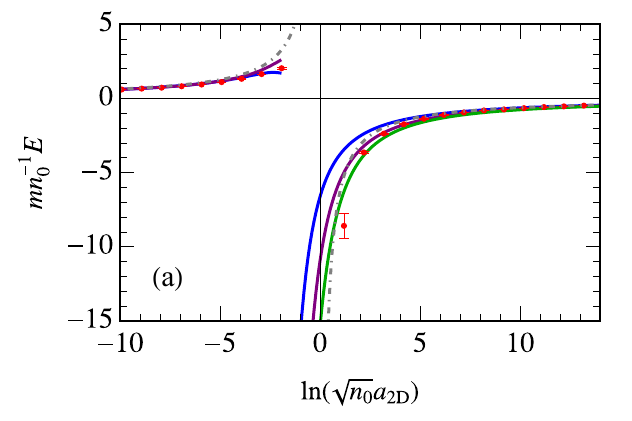}
    \label{fig:energy 1}
    \end{minipage}
    \begin{minipage}{0.48\textwidth}
    \centering
    \includegraphics[width=\textwidth]{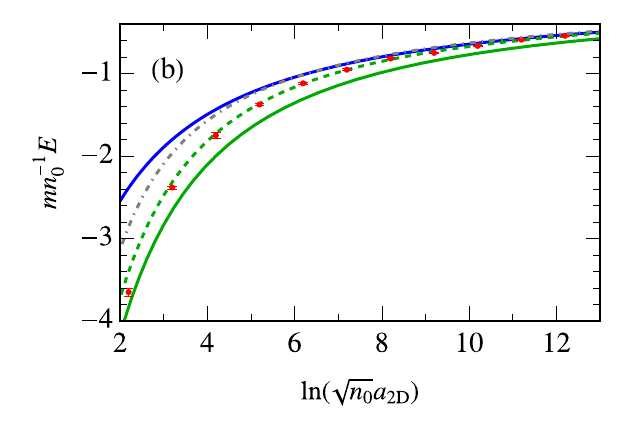}
    \label{fig:energy 2}
    \end{minipage}
    \caption{Energy of the Bose polaron as a function of the coupling strength $\ln({\sqrt{n_{0}}a_{\mathrm{2D}}})$. (a) The blue, purple, and green lines show the results obtained from two-, three-, and four-body correlations, respectively, for the ideal BEC with $\mu = 0$. The red dots and grey dash-dotted line show the results obtained from QMC~\cite{Ardila2020} and perturbation theory~\cite{Pastukhov2018b}, respectively, with $\mu = 0.136 n_0/m$. (b) A closeup of the weak-coupling attractive polaron in (a), where the additional green dashed line shows the result obtained from four-body correlations for a weakly interacting BEC with $\mu = 0.136 n_0/m$. The variational energy  obtained from two-body correlations at this $\mu$ is indistinguishable from the corresponding result for $\mu=0$ on this scale.}
    \label{fig:energy}
\end{figure}

In Fig.~\ref{fig:energy}(a), we display the polaron energy with $\mu = 0$ (corresponding to taking the limit of a non-interacting BEC, i.e., $a_{\mathrm B}\to0^+$). Similarly to the 3D polaron, we observe two branches: the ground-state attractive branch and the metastable repulsive branch. As discussed in Appendix~\ref{appx: couped integral equations}, the polaron energy in the case of two- and three-body correlations is obtained by solving the coupled integral equations in Eq.~\eqref{eq:coupled integral equations}, while the attractive polaron energy incorporating four-body correlations is instead obtained from the reduced coupled equations in Eq.~\eqref{eq:reduced coupled equations}. These two approaches are completely equivalent.

We first consider the attractive branch. In the weak-coupling regime where the size of the bound state greatly exceeds the interparticle spacing, $\ln({\sqrt{n_{0}}a_{\text{2D}}}) \gg 1$, we observe that our variational results from two-, three- and four-body correlations are in excellent agreement with both QMC~\cite{Ardila2020} and perturbation theory~\cite{Pastukhov2018b} (which were both calculated at $\mu=0.136 n_0/m$). Our variational results begin to deviate with increasing coupling strength, as multi-body correlations become more important. However, even in the strong-coupling regime where $a_{\mathrm{2D}}\sim n_0^{-1/2}$, our variational results from four-body correlations are still in good agreement with QMC, which is reminiscent of the case for the 3D Bose polaron \cite{Yoshida2018PRX,Field2020,Levinsen2021}.

While at a qualitative level all methods agree well for weak to moderate interaction strengths, we can gain further insight by taking a closer look at the results in this regime, as shown in Fig.~\ref{fig:energy}(b). First, we see that perturbation theory is well reproduced already by the variational ansatz limited to two-body correlations, which we find are nearly completely independent of $\mu$. Incorporating up to four-body correlations, we see that the Bose gas chemical potential matters more in this case. This is to be expected since including higher-order correlations is necessary in order to capture the correction due to Lee-Huang-Yang-type quantum fluctuations, as demonstrated in the 3D case~\cite{Christensen2015,Levinsen2015}. Especially, we observe that the inclusion of four-body correlations and a finite $\mu$ allows us to almost perfectly reproduce the QMC results of Ref.~\cite{Ardila2020} in this regime, highlighting the importance of incorporating the chemical potential. On the other hand, the polaron energy is dominated by few-body correlations in the strong-coupling regime and it is currently an open question whether one would need to include the interboson repulsion beyond the Bogoliubov approximation to obtain exact agreement with the last point of QMC for the attractive branch. Certainly, this repulsion will eventually need to be included for very tightly bound attractive polarons.

We now turn to the repulsive branch energy, as shown in Fig.~\ref{fig:energy}(a), where we obtain the energy by finding the peak of the spectral function at positive energy. While the line shape of the spectral response plays a small role, we find that the repulsive polaron energy is quite insensitive to the Fourier broadening. Similarly to the attractive branch, we see that our variational results from two- and three-body correlations have excellent agreement with QMC and perturbation theory in the weak-coupling regime, $\ln({\sqrt{n_{0}}a_{\text{2D}}}) \ll -1$. Towards the strong-coupling regime, the polaron energy becomes comparable to the lifetime of the polaron, indicating that the repulsive polaron ceases to correspond to a well-defined quasiparticle. This results in the small difference in the corresponding repulsive polaron energies around $\ln({\sqrt{n_{0}}a_{\text{2D}}}) \simeq -1$.

\begin{figure}
    \centering
    \begin{minipage}{0.5\textwidth}
    \centering
    \includegraphics[width=0.95\textwidth]{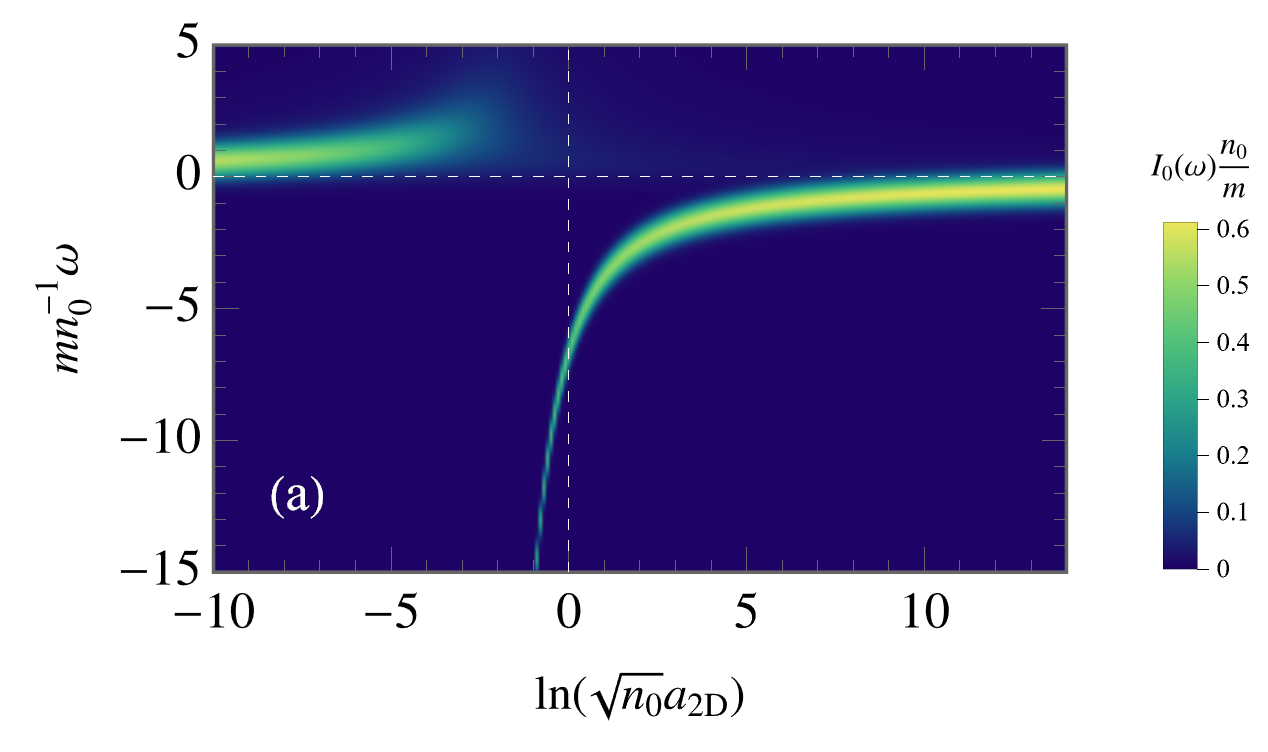}
    \label{fig:spectral response 2 body}
    \end{minipage}
    \begin{minipage}{0.5\textwidth}
    \centering
    \includegraphics[width=0.95\textwidth]{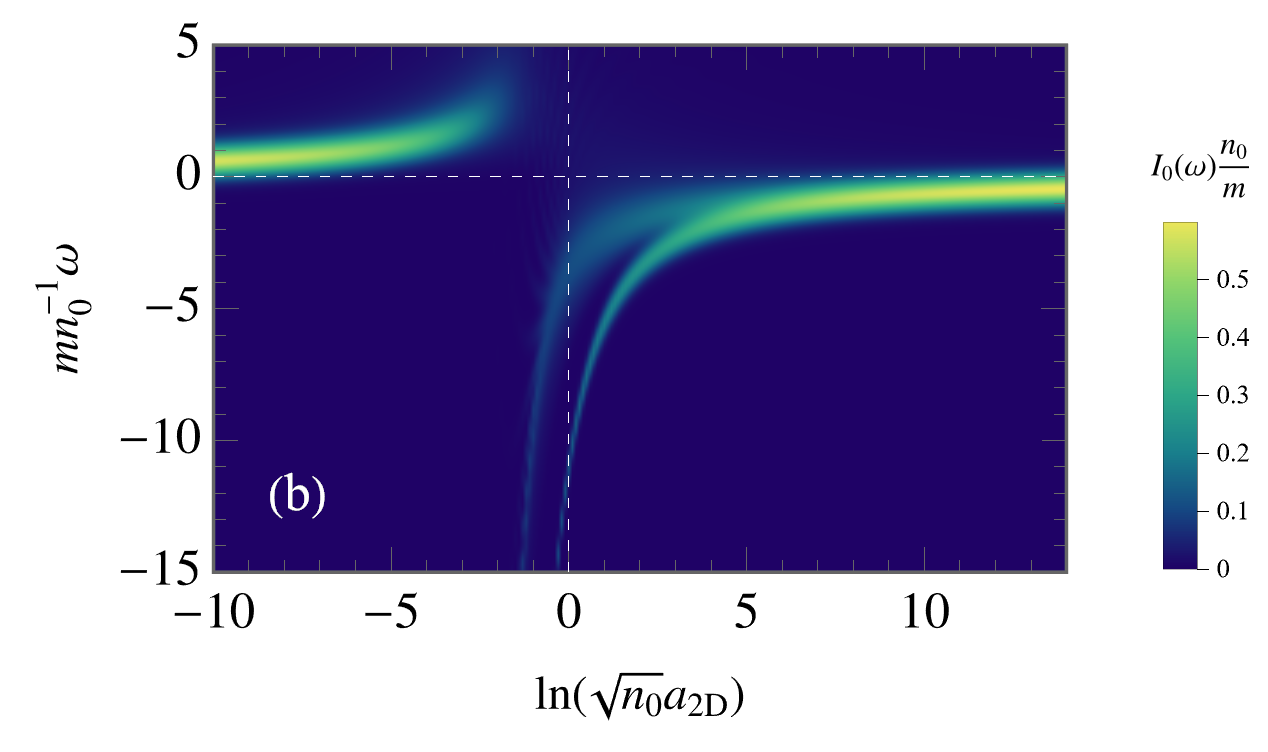}
    \label{fig:spectral response 3 body}
    \end{minipage}
    \caption{Spectral response of the impurity in the BEC obtained from (a) two-body correlations and (b) three-body correlations with $\mu = 0$. The spectral response for a weakly interacting BEC with $\mu = 0.136 n_0/m$ is not included as it is almost identical to that for the ideal BEC, with slight broadening in the excited attractive branch for three-body correlations.}
    \label{fig:spectral response}
\end{figure}

In Fig.~\ref{fig:spectral response}, we show the spectral response obtained from (a) two-body correlations and (b) three-body correlations. In both panels we observe that the repulsive branch gets broadened towards the strong-coupling regime, indicating a higher uncertainty of the polaron energy and an associated shorter lifetime. Thus, decay processes including many-body dephasing~\cite{Adlong2020} and relaxation into the lower-lying continuum of states dominate in this regime. In (b), we also observe the emergence of an additional excited state of the attractive polaron. While the lower attractive branch corresponds to a single eigenstate involving the impurity dressed by up to two excitations of the medium, the upper attractive branch corresponds to a continuum of states involving the impurity dressed by at most one excitation, moving with respect to another excitation. Indeed, the upper attractive branch starts from the ground-state energy of two-body correlations; however due to the small density of states at small relative momentum, the peak appears slightly shifted compared with the peak in panel (a).

It is an interesting and currently open question how the exact spectral function of the Bose polaron behaves in the strong-coupling regime. While we have seen that the repulsive polaron changes little upon adding a second excitation of the medium (and therefore it is unlikely to change much if one could do an exact calculation), we believe that each extra excitation included in our ansatz will result in an extra attractive polaron branch, similar to what is seen in the difference between Fig.~\ref{fig:spectral response}(a) and (b). In the case of an infinitely heavy impurity in a 3D non-interacting BEC, it has been shown analytically~\cite{Drescher2021} that the attractive polaron spectrum becomes very smooth at unitarity, and features multiple branches for positive scattering lengths when there is a bound state. Since in 2D there is always a bound state, it is therefore reasonable to assume that there will be multiple branches also in the exact solution, although the precise role of the boson-boson repulsion would need to be elucidated.

\begin{figure}[htp]
    \centering
    \begin{minipage}{0.48\textwidth}
    \centering
    \includegraphics[width=\textwidth]{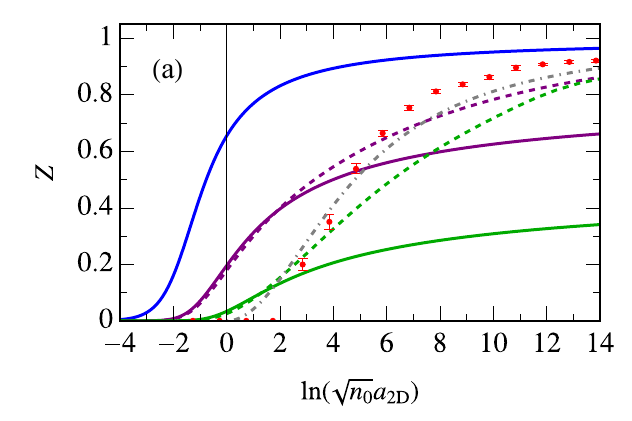}
    \label{fig:attractive residue}
    \end{minipage}
    \begin{minipage}{0.48\textwidth}
    \centering
    \includegraphics[width=\textwidth]{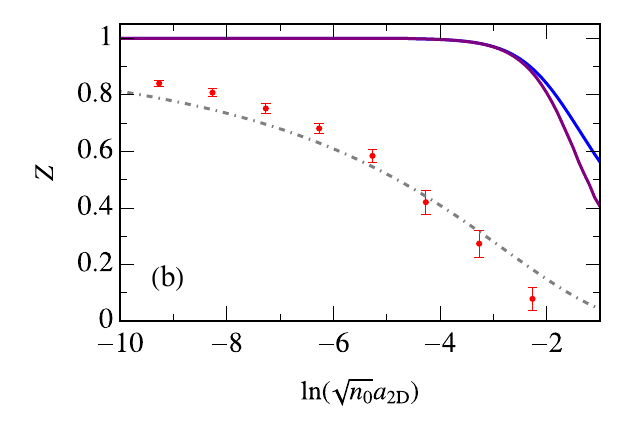}
    \label{fig:repulsive residue}
    \end{minipage}
    \caption{Residue of (a) the attractive Bose polaron and (b) the repulsive Bose polaron as a function of the coupling strength $\ln({\sqrt{n_{0}}a_{\text{2D}}})$. The blue, purple, and green solid lines show the results obtained from two-, three, and four-body correlations, respectively, for the ideal BEC with $\mu = 0$. The red dots and grey dash-dotted line show the results obtained from QMC~\cite{Ardila2020} and perturbation theory~\cite{Pastukhov2018b}, respectively, with $\mu = 0.136 n_0/m$. The purple and green dashed lines show the results from three and four-body correlations, respectively, for a weakly interacting BEC with $\mu = 0.136 n_0/m$ (for the attractive polaron obtained from two-body correlations and for the repulsive polaron, we do not see a visible difference between $\mu=0$ and $\mu=0.136 n_0/m$ on this scale).}
    \label{fig:residue}
\end{figure}

\subsection{Residue}
\label{subsec:residue}
The residue $Z$ quantifies the squared overlap of the interacting state with the non-interacting state. Thus, for the attractive polaron which consists of only a single state, the residue reads:
\begin{equation}
    Z = |\alpha_{0}|^2.
\end{equation}
For the repulsive branch which corresponds to a continuum of states, we define the residue as the sum of all states with positive energy, $Z=\sum_{j, E_j>0}|\alpha_0^{(j)}|^2$. We find that this definition matches well with previous calculations for a single Bogoliubov excitation~\cite{Castillo2023}.

In Fig.~\ref{fig:residue}(a), we display the attractive polaron residue obtained from our variational approach including two-, three-, and four-body correlations. For the case of the ideal BEC, we observe a clear difference between our results in the weak-coupling regime, $\ln(\sqrt{n_{0}}a_{\text{2D}}) \gg 1$. The residue in the case of two-body correlations remains close to 1, while in the case of three- and four-body correlations, it appears significantly lowered. There are two reasons for this: First, the existence of bound few-body states
enhance impurity dressing by the excitations of the medium, and second the infinite compressibility of the ideal BEC means that the impurity can excite an arbitrary number of low-energy excitations, thus leading to the so-called orthogonality catastrophe~\cite{Yoshida2018PRX,Mistakidis2019,Guenther2021}. 
While the latter effect cannot be captured within our variational approach (being limited to few-body correlations), it still suppresses the residue when considering three- and four-body correlations. 

Unlike the case of the polaron energy, we find that the residue strongly depends on the medium chemical potential, which changes the compressibility of the BEC. Indeed, upon including a small chemical potential, we observe in the weak-coupling regime that the residue obtained from three- and four-body correlations approaches values closer to those from two-body correlations, QMC, and perturbation theory. Most notably, in the weak-coupling regime, the three- and four-body correlations now match, indicating that they well describe the leading-order perturbative corrections to the mean-field result, similarly to the 3D case~\cite{Christensen2015,Levinsen2015}. This is in contrast to the QMC results, which appear to overestimate the residue in this limit. On the other hand, in the strong-coupling regime, the residue is almost completely unaffected by the chemical potential of the BEC and is instead determined by strong few-body correlations. These will, in turn, be affected by the short-range boson-boson repulsion in a manner beyond the Bogoliubov approximation.

For the repulsive polaron shown in Fig.~\ref{fig:residue}(b), we find that the result is nearly completely independent of the medium chemical potential and is insensitive to the number of excitations, unlike the attractive case. However, we see that our variational results strongly deviate from QMC and perturbation theory. 
This suggests that the metastability of the repulsive polaron might play a strong role in the behavior of the spectral weight, since this is not accounted for in the QMC which assumes that the repulsive polaron is an excited eigenstate. 
We also note that our calculations yield attractive and repulsive polaron residues that nearly sum to 1, as expected, while QMC finds that substantial spectral weight shifts elsewhere in the spectrum. This discrepancy warrants additional future studies.

\subsection{Effective Mass}
\label{subsec:effective mass}
Thus far, we have considered states with zero total momentum, which is relevant for the spectrum at zero temperature. However, our approach also contains information about the behavior at small momentum $\p$ and thus the dynamical response of the system. 
For the case of the attractive polaron, where there is a well-defined quasiparticle, the polaron energy can be expanded as:
\begin{equation}
    E(p) = E(0) + \frac{p^2}{2m^{\ast}} + \mathcal{O}(p^4) \label{eq:energy dispersion at small momentum},
\end{equation}
where $m^{\ast}$ is the polaron effective mass. This modified mass arises from the interactions between the impurity and the medium, which affects the mobility of the impurity. We calculate the polaron effective mass using our variational ansatz, as outlined in Appendix~\ref{appx: effective mass}.

Figure~\ref{fig:effective mass} shows the effective mass of the attractive polaron obtained from two- and three-body correlations. Similarly to the residue of the attractive polaron in the weak-coupling regime, the effective mass obtained from three-body correlations in the case of the ideal BEC appears substantially larger than the results from two-body correlations. This signals enhanced impurity dressing by excitations of the medium, which can be compensated by including weak interactions in the BEC (corresponding to a small chemical potential). The latter is seen to match perturbation theory in the weak-coupling regime and qualitatively agrees with QMC for $\ln(\sqrt{n_0}a_{\mathrm 2D})\gtrsim 4$. 
Towards the strong-coupling limit, the effective mass in our variational ansatz converges to $m^{\ast} = (N+1)m$, where $N$ is the number of bosons included in the evaluations, and thus cannot capture the dressing by many bosons seen in the QMC. However, while our ansatz does not describe the ground state in this regime, our calculation might instead effectively describe an excited metastable polaron with a spectral weight greater than that of the true ground-state polaron [whose residue vanishes in this regime according to QMC, as shown in Fig.~\ref{fig:residue}(a)].

\begin{figure}
    \centering
    \includegraphics[width=0.5\textwidth]{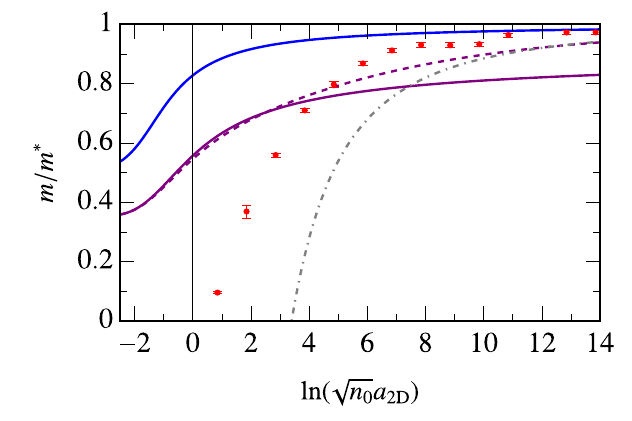}
    \caption{Effective mass $m^*$ of the attractive Bose polaron as a function of the coupling strength $\ln({\sqrt{n_{0}}a_{\text{2D}}})$. The blue and purple lines show the results obtained from two- and three-body correlations, respectively, for the ideal BEC with $\mu = 0$. The red dots and grey dash-dotted line show the results obtained from QMC~\cite{Ardila2020} and perturbation theory~\cite{Pastukhov2018b}, respectively, with $\mu = 0.136 n_0/m$. The purple dashed line shows the results from three-body correlations for a weakly interacting BEC with $\mu = 0.136 n_0/m$ (the corresponding result for two-body correlations at this $\mu$ is not included as it is indistinguishable from the result for $\mu=0$ on this scale).}
    \label{fig:effective mass}
\end{figure}
 
\section{Conclusions}
\label{sec:conclusions}
In conclusion, we have extensively examined the properties of the 2D Bose polaron using a variational approach that incorporates up to three excitations of the medium. We have shown that considering four-body correlations and a weak BEC chemical potential is crucial to achieve agreement with the results obtained from QMC calculations. Particularly in the weak-coupling regime, our variational approach, accounting for four-body correlations in the presence of a weakly interacting BEC, exhibits excellent agreement with QMC results for the polaron energy. However, as we approach the strong-coupling regime, discrepancies arise between the energy values obtained from our variational approach and QMC due to strong few-body correlations.
Furthermore, we have found that the residue and effective mass of the attractive polaron strongly depend on the level of correlations we incorporate in our ansatz, as well as the interactions within the BEC. On the other hand, the residue of the repulsive branch is remarkably insensitive to the number of Bogoliubov excitations and appears to be strongly influenced by the metastable nature of the repulsive polaron. 

Our results can potentially be probed in ultracold atomic gas experiments, similarly to how the 2D Fermi polaron was previously investigated~\cite{Koschorreck2012,Ong2015,Oppong2019}. The Bose polaron may also be investigated in 2D semiconductor microcavities with exciton polaritons (bound electron-hole pairs strongly coupled to light). Here, exciton-polariton ``impurities'' in a given circular polarization are dressed by a medium of exciton polaritons in the opposite polarization~\cite{Levinsen2019}. Indeed, this intriguing extension of traditional polaron physics has recently been realized in a \ch{MoSe2} monolayer~\cite{Tan2023}. While the interplay between drive/dissipation and polaron physics in these systems is still an open question, these emerging experimental platforms offer exciting opportunities to probe and further investigate the properties and behavior of the Bose polaron.

\acknowledgments 
We acknowledge insightful discussions with Luis A.~Pe\~na Ardila and we thank him for sharing the QMC data from Ref.~\cite{Ardila2020}. We also acknowledge discussions with Arturo Camacho-Guardian. We acknowledge support from the Australian Research Council
Centre of Excellence in Future Low-Energy Electronics Technologies
(CE170100039).  MMP and JL are also supported through the Australian Research
Council Future Fellowships FT160100244 and FT200100619, respectively. 

\onecolumngrid
\appendix
\section{Coupled integral equations}
\label{appx: couped integral equations}
The stationary condition $\bra{\partial\Psi}(\hat{H}-E)\ket{\Psi} = 0$ with respect to the variational parameters $\alpha^{\ast}_{0}$, $\alpha^{\ast}_{\k}$, $\alpha^{\ast}_{\kone\ktwo}$, $\alpha^{\ast}_{\kone\ktwo\kthree}$, $\gamma^{\ast}_{0}$, $\gamma^{\ast}_{\k}$, and $\gamma^{\ast}_{\kone\ktwo}$ yields seven coupled equations~\cite{Levinsen2015,Yoshida2018PRX}:
\begin{align}
    E\alpha_{0} &= g\sqrt{n_{0}}\gamma_{0} - g\sum_{\k}v_{\k}\gamma_{\k}, \notag
    \\
    (E-\ek-E_{\k})\alpha_{\k} &= gu_{\k}\gamma_{0} + g\sqrt{n_0}\gamma_{\k} -g\sum_{\q}\gamma_{\k\q}v_{\q}, \notag 
    \\
    (E-E_{\kone\ktwo})\alpha_{\kone\ktwo} &= g(\gamma_{\kone}u_{\ktwo} + \gamma_{\ktwo}u_{\kone}) + g\sqrt{n_0}\gamma_{\kone\ktwo}, \notag
    \\
    (E-E_{\kone\ktwo\kthree})\alpha_{\kone\ktwo\kthree} &= g(\gamma_{\kone\ktwo}u_{\kthree} + \gamma_{\ktwo\kthree}u_{\kone} + \gamma_{\kone\kthree}u_{\ktwo}), \notag
    \\
    (E-\nu_{0})\gamma_{0} &= g\sqrt{n_{0}}\alpha_{0} + g\sum_{\k}u_{\k}\alpha_{\k}, \notag
    \\
    (E-\ek^{d}-\nu_{0}-E_{\k})\gamma_{\k} &= g\sqrt{n_0}\alpha_{\k} - gv_{\k}\alpha_{0} + g\sum_{\k'}u_{\k'}\alpha_{\k\k'}, \notag
    \\
    (E-\epsilon^{d}_{\kone+\ktwo}-\nu_{0}-E_{\kone}-E_{\ktwo})\gamma_{\kone\ktwo} &= g\sqrt{n_0}\alpha_{\kone\ktwo} - g(\alpha_{\kone}v_{\ktwo}+\alpha_{\ktwo}u_{\kone}) + g\sum_{\kthree}\alpha_{\kone\ktwo\kthree}u_{\kthree},
    \label{eq:coupled integral equations}
\end{align}
where $E_{\kone\ktwo} \equiv E_{\kone}+E_{\ktwo}+\epsilon_{\kone+\ktwo}$ and $E_{\kone\ktwo\kthree} \equiv E_{\kone}+E_{\ktwo}+E_{\kthree}+\epsilon_{\kone+\ktwo+\kthree}$.

Using the first four equations to remove the $\alpha$ parameters, we obtain the reduced coupled equations:
\begin{align}
    \mathcal{T}^{-1}(E,\mathbf{0})\gamma_{0} &= \frac{n_{0}}{E}\gamma_{0} + \sqrt{n_{0}}\sum_{\k} \left( \frac{u_{\k}\gamma_{\k}}{E-\ek-E_{\k}} - \frac{v_{\k}\gamma_{\k}}{E} \right) - \sum_{\k\q}\frac{u_{\k}v_{\q}\gamma_{\k\q}}{E - \ek - E_{\k}}, \notag
    \\
    \mathcal{T}^{-1}(E - E_\k,\k)\gamma_{\k} &= \sqrt{n_{0}} \left( \frac{u_{\k}}{E-\ek-E_{\k}} - \frac{v_{\k}}{E} \right)\gamma_{0} + \frac{n_{0}}{E-\ek-E_{\k}}\gamma_{\k} + \sum_{\q} \left( \frac{u_{\k}u_{\q}\gamma_{\q}}{E-E_{\k\q}}+\frac{v_{\k}v_{\q}\gamma_{\q}}{E} \right) \notag \\
    &\quad + \sqrt{n_0}\sum_{\q}\left( \frac{u_{\q}\gamma_{\k\q}}{E-E_{\k\q}} - \frac{v_{\q}\gamma_{\k\q}}{E-\ek-E_{\k}} \right), \notag
    \\
    \mathcal{T}^{-1}(E - E_{\kone} - E_{\ktwo} , \kone + \ktwo)\gamma_{\kone\ktwo} &= \frac{n_0}{E-E_{\kone\ktwo}}\gamma_{\kone\ktwo} + \biggl[ \biggl\{ \sqrt{n_0} \left(\frac{u_{\kone}\gamma_{\ktwo}}{E-E_{\kone\ktwo}} - \frac{v_{\kone}\gamma_{\ktwo}}{E-\epsilon_{\ktwo}-E_{\ktwo}}\right) - \frac{u_{\kone}v_{\ktwo}\gamma_0}{E-\epsilon_{\kone}-E_{\kone}} \notag \\
    &\quad + \sum_{\q} \left( \frac{u_{\kone}u_{\q}\gamma_{\ktwo\q}}{E-E_{\kone\ktwo\q}} + \frac{v_{\kone}v_{\q}\gamma_{\ktwo\q}}{E-\epsilon_{\ktwo}-E_{\ktwo}} \right) \biggl\} + (\kone \leftrightarrow \ktwo) \biggl],
    \label{eq:reduced coupled equations}
\end{align}
where we have defined the medium $T$ matrix $\mathcal{T}(E,\k)$:
\begin{align}
    \mathcal{T}^{-1}(E,\k) &= \frac{m^2R^2_{\text{2D}}}{4\pi}\left( E + \eb - \epsilon^{d}_{\k} \right) - \frac{m}{4\pi}\ln{\left( \frac{-E}{\eb} \right)} - \sum_{\q} \left( \frac{u^2_{\q}}{E - E_{\q}-\epsilon_{\k+\q}} + \frac{1}{2\epsilon_{\q} - E} \right).
\end{align}

We can obtain the quasiparticle properties from either of Eqs.~\eqref{eq:coupled integral equations} or \eqref{eq:reduced coupled equations}. In our present work, the attractive and repulsive polaron energy and residue for two- and three-body correlations are obtained from Eq.~\eqref{eq:coupled integral equations}, while the attractive polaron energy and residue for four-body correlations are obtained from Eq.~\eqref{eq:reduced coupled equations}. The attractive polaron energy and residue in the case of two- and three-body correlations can also be obtained from Eq.~\eqref{eq:reduced coupled equations}, and we have checked that the two cases yield exactly the same results in the limit of $\Lambda\to\infty$ and $\nu_0\to\infty$.

\section{Residue}
To evaluate the residue of the attractive polaron from Eq.~\eqref{eq:reduced coupled equations}, we define:
\begin{equation}
    \xi_0 = \frac{g^{2}\sum_{\k}u_{\k}\alpha_{\k}}{E-\nu_{0}}, \quad \xi_{\k} = \frac{g^{2}\sum_{\k'}u_{\k'}\alpha_{\k\k'}}{E-\ek^{d}-\nu_{0}-E_{\k}}, \quad \xi_{\kone\ktwo} = \frac{g^{2}\sum_{\kthree}\alpha_{\kone\ktwo\kthree}u_{\kthree}}{E-\epsilon^{d}_{\kone+\ktwo}-\nu_{0}-E_{\kone}-E_{\ktwo}},
\end{equation}
which remain finite in the limit $\nu_0 \rightarrow \infty$. Assuming that the variational parameters are normalized according to $\bra{\Psi}\ket{\Psi} = 1$, the residue of the attractive polaron is given by $Z = |\alpha_{0}|^2$. Carrying out the renormalization procedure, this takes the form:
\begin{align}
    Z &= \left( \frac{\sqrt{n_0}\xi_0}{E} - \sum_{\k}\frac{v_{\k}\xi_{\k}}{E} \right)^2 \Biggl[ \left( \frac{\sqrt{n_0}\xi_0}{E} - \sum_{\k}\frac{v_{\k}\xi_{\k}}{E} \right)^2 + \sum_{\k}\left( \frac{u_{\k}\xi_0+\sqrt{n_0}\xi_{\k}-\sum_{\q}v_{\q}\xi_{\k\q}}{E-\ek-E_{\k}} \right)^2 \notag \\
    &\quad + \sum_{\kone\ktwo}\left( \frac{u_{\kone}\xi_{\ktwo}+u_{\ktwo}\xi_{\kone}+\sqrt{n_0}\xi_{\kone\ktwo}}{E-E_{\kone\ktwo}}\right)^2 + \sum_{\kone\ktwo\kthree}\left( \frac{u_{\kone}\xi_{\ktwo\kthree}+u_{\ktwo}\xi_{\kthree\kone}+u_{\kthree}\xi_{\kone\ktwo}}{E-E_{\kone\ktwo\kthree}} \right)^2 + \frac{\xi^{2}_{0}}{g^2} + \sum_{\k}\frac{\xi^{2}_{\k}}{g^2} + \sum_{\kone\ktwo}\frac{\xi^{2}_{\kone\ktwo}}{g^2} \Biggl]^{-1}.
\end{align}

\section{Effective mass}
\label{appx: effective mass}
To evaluate the polaron effective mass, we extend our variational ansatz to include a finite momentum $\p$:
\begin{equation}
    \ket{\Psi^{(\p)}} = \left(\alpha^{(\p)}_{0}c^{\dagger}_{\p} + \sum_{\k}\alpha^{(\p)}_{\k}c^{\dagger}_{\p-\k}\beta^{\dagger}_{\k} + \frac{1}{2}\sum_{\kone\ktwo}\alpha^{(\p)}_{\kone\ktwo}c^{\dagger}_{\p-\kone-\ktwo}\beta^{\dagger}_{\kone}\beta^{\dagger}_{\ktwo} + \gamma^{(\p)}_{0}d^{\dagger}_{\p} + \sum_{\k}\gamma^{(\p)}_{\k}d^{\dagger}_{\p-\k}\beta^{\dagger}_{\k} \right)\ket{\Phi},
\end{equation}
where we consider the terms up to three-body correlations for simplicity. The impurity momentum breaks rotational symmetry and the variational parameters now depend on the angle between the total momentum and each wavevector of the Bogoliubov excitations. Taking the stationary condition, we obtain the modified coupled integral equations:
\begin{align}
    (E(p)-\ep)\alpha^{(\p)}_{0} &= g\sqrt{n_{0}}\gamma^{(\p)}_{0} - g\sum_{\k}v_{\k}\gamma^{(\p)}_{\k}, \notag \\
    (E(p)-\epsilon_{\k-\p}-E_{\k})\alpha^{(\p)}_{\k} &= gu_{\k}\gamma^{(\p)}_{0} + g\sqrt{n_0}\gamma^{(\p)}_{\k}, \notag \\
    (E(p)-E_{\kone}-E_{\ktwo}-\epsilon_{\kone+\ktwo-\p})\alpha^{(\p)}_{\kone\ktwo} &= g(\gamma^{(\p)}_{\kone}u_{\ktwo} + \gamma^{(\p)}_{\ktwo}u_{\kone}), \notag \\
    (E(p)-\ep^{d}-\nu_{0})\gamma^{(\p)}_{0} &= g\sqrt{n_{0}}\alpha^{(\p)}_{0} + g\sum_{\k}u_{\k}\alpha^{(\p)}_{\k}, \notag \\
    (E(p)-\epsilon^{d}_{\k-\p}-\nu_{0}-E_{\k})\gamma^{(\p)}_{\k} &= g\sqrt{n_0}\alpha^{(\p)}_{\k} - gv_{\k}\alpha^{(\p)}_{0} + g\sum_{\k'}u_{\k'}\alpha^{(\p)}_{\k\k'},
    \label{eq:coupled integral equations with finite momentum}
\end{align}
where $E(p)$ is the energy dispersion of the Bose polaron with a total momentum $\p$.

For a small momentum $\p$, the energy dispersion can be expanded as:
\begin{equation}
    E(p) = E(0) + \frac{p^2}{2m^{\ast}} + \mathcal{O}(p^4) \label{eq:energy dispersion at small momentum 2},
\end{equation}
where $m^{\ast}$ is the polaron effective mass. In principle, the effective mass can be obtained by solving Eqs.~\eqref{eq:coupled integral equations with finite momentum} and \eqref{eq:energy dispersion at small momentum 2} at various momenta. However, this procedure is numerically cumbersome due to the absence of rotational symmetry. We avoid this problem by employing the perturbative approach introduced in Ref.~\cite{Trefzger2012} (see also Ref.~\cite{Yoshida2018PRX}). Since the coupled integral equations in Eq.~\eqref{eq:coupled integral equations with finite momentum} are all linear, they can be expressed as the eigenvalue problem:
\begin{equation}
    \Xh(E(p),\p)\ket{\eta} = 0,
\end{equation}
where $\ket{\eta}$ denotes the ordered set of variational parameters  $\alpha^{(\p)}_{0}$, $\alpha^{(\p)}_{\k}$, $\alpha^{(\p)}_{\kone\ktwo}$, $\gamma^{(\p)}_{0}$, and $\gamma^{(\p)}_{\k}$. Without loss of generality, we assume that the momentum $\p$ is oriented along the $x$ axis. The energy dispersion of the ground state polaron corresponds to the zero-crossing of the lowest (or highest, depending on the sign convention of the operator $\Xh$) eigenvalue of $\Xh$. Assuming that the momentum $\p$ is small, the Taylor series of $\Xh$ up to $\mathcal{O}(p^2)$ reads:
\begin{equation}
     \Xh(E(p),\p) = \Xh_{0} + p\Xh_{p} + \frac{p^2}{2}\left(\Xh_{pp} + \frac{1}{m^\ast}\Xh_{E} \right) + \mathcal{O}(p^3), \label{eq:x-expansion}
\end{equation}
where we have defined operators: 
\begin{equation}
    \Xh_{0} = \Xh(E(0),0), \quad \Xh_{p} = \left.\frac{\partial\Xh}{\partial p}
    \right|_{\tiny\subalign{p&=0 \\ E&=E(0)}}, \quad \Xh_{pp} = \left.\frac{\partial^2\Xh}{\partial p^2}\right|_{\tiny\subalign{p&=0 \\ E&=E(0)}}, \quad \Xh_{E} = \left.\frac{\partial\Xh}{\partial E}\right|_{\tiny\subalign{p&=0 \\ E&=E(0)}}.
\end{equation}
Treating the operators proportional to $p$ in Eq.~\eqref{eq:x-expansion} as small perturbations to $\Xh_{0}$, i.e., the coupled integral equations for a $\p = 0$ impurity, the lowest eigenvalue $\lambda(p)$ of $\Xh(E(p),\p)$ can be expressed as:
\begin{equation}
    \lambda(p) = \lambda_{0} + \lambda_{1}p + \lambda_{2}p^2 + \mathcal{O}(p^3) \label{eq:lowest eigenvalue}
\end{equation}
The first- and second-order corrections $\lambda_{i}$ are calculated from perturbation theory:
\begin{align}
    \lambda_{1} &= \bra{\eta^{(0)}}\Xh_{p}\ket{\eta^{(0)}} = 0, \label{eq:first coefficient} \\
    \lambda_{2} &= \frac{1}{2}\bra{\eta^{(0)}}\left(\Xh_{pp}+\frac{1}{m^{\ast}}\Xh_{E}\right)\ket{\eta^{(0)}} + \sum_{i>0}\frac{\left|\bra{\eta^{(i)}}\Xh_{p}\ket{\eta^{(0)}}\right|^{2}}{\lambda^{(0)}_{0}-\lambda^{(i)}_{0}}, \label{eq:second coefficient}
\end{align}
where $\lambda^{(i)}_{0}$ and $\ket{\eta^{(i)}}$ denote the $i$th eigenvalue and its corresponding eigenvector of $\Xh_{0}$, and $i=0$ indicates the ground state. Recalling that the ground state energies of the polaron with $\p = 0$ and $\p \neq 0$ are given, respectively, by the zero-crossing of the lowest eigenvalues of $\Xh_{0}$ and ${\Xh(E(p),\p)}$, we can set  $\lambda^{(0)}_{0}$ = 0 and $\lambda{(p)} = 0$. In calculating the first-order correction $\lambda_{1} = 0$, we have taken the $s$-wave average of the expectation value. Thus, we conclude that $\lambda_{2} = 0$ and obtain the inverse effective mass:
\begin{align}
    \frac{1}{m^{\ast}} &= \bra{\eta^{(0)}}\Xh_{E}\ket{\eta^{(0)}}^{-1}\left[\sum_{i>0}\frac{2}{\lambda^{(i)}_{0}}\left|\bra{\eta^{(i)}}\Xh_{p}\ket{\eta^{(0)}}\right|^{2} - \bra{\eta^{(0)}}\Xh_{pp}\ket{\eta^{(0)}} \right] \notag \\
    &= \bra{\eta^{(0)}}\Xh_{E}\ket{\eta^{(0)}}^{-1}\left[2\bra{\eta^{(0)}}\Xh_{p}\hat{Q}\Xh^{-1}_{0}\hat{Q}\Xh_{p}\ket{\eta^{(0)}} - \bra{\eta^{(0)}}\Xh_{pp}\ket{\eta^{(0)}}\right],
\end{align}
where $\hat{Q} = \mathbb{1} - \ket{\eta^{(0)}}\bra{\eta^{(0)}}$.

\twocolumngrid

\bibliography{cold-atoms}

\end{document}